\documentstyle[preprint,revtex]{aps}
\begin{document}
\draft
\preprint{}
\begin{title}
Thermal history of the string universe
\end{title}
\author{Hideyuki C$ \hat {\hbox{A}}$TEAU and Kosuke SUMIYOSHI }
\begin{instit}
Department of Physics   \\
 Tokyo Metropolitan University \\
 1-1 Minami-Osawa, Hachioji, \\
 Tokyo 192-03  Japan
\end{instit}
\begin{abstract}
Thermal history of the string universe
based on the Brandenberger and Vafa's scenario
is examined.   The analysis thereby provides  a
theoretical foundation of the string universe scenario.
Especially the picture of the initial oscillating phase is shown to be
 natural from the thermodynamical point of view.
A precise description is also given of
the transition process from the stringy phase
to the radiation dominated phase.

\end{abstract}

\narrowtext
\section{INTRODUCTION}
\label{sec:intro}

Since the Green and Schwarz's anomaly cancellation
\cite{GSW}
 proved the importance of
the super string theory(SST), detailed studies of it have been done.
One unfortunate feature of SST
 is the fact that its typical energy scale is
$ 10^{19}$ GeV which is far beyond our experimental access.
However this does not necessarily imply that SST allows no experimental
test.
The most important feature of SST is that it unifies the theories
 of matter and
the gravity. This  implies
  that SST in principle has an ability to determine
the structure of  space time in which the strings themselves live.
If SST is  the true theory of the whole universe, it is conceivable that
SST has left some relics in our universe observable even today.
In fact the presently observed isotropic, uniform
 and almost flat universe must have
been  determined by  SST.
{}From this point of view the string cosmology has been studied by
 some authors \cite{Witten,Ah,BV,MM,TV}.

Brandenberger and Vafa \cite{BV} proposed
 an interesting scenario of the string
cosmology.  The starting point of their scenario
is  the Heterotic string theory in the space of nine
dimensional torus universe of the Plankian size and  a time dimension
$T^9\times R$.
They argued that this small universe was oscillating in some period and
eventually three dimensions out of nine began to expand resulting in the
 present large universe.

In order to get deeper theoretical understanding of their scenario,
 we perform a detailed thermodynamical analysis of it in this paper.
Our strategy in this paper is to use the microcanonical formalizm to
follow an entire thermal history.  So far
 several authors have employed the
microcanonical formalism to examine the thermodynamical functions
of the string gas.  However the relation of their results to the
thermal history of the string universe seems to be  unclear  to us.

In order to clear up these situations in this paper we give a
concrete framework by which we
 can follow the thermal history of the string
universe
using the thermodynamical functions of the microcanonical formalism.
According to this framework
and based on some assumptions such as the local thermal equilibrium
 and others which we will
state precisely later,
we will determine the thermal history of the
 nine dimensional torus universe of Brandenberger and Vafa  as  follows.

In the initial epoch during which the torus universe is oscillating,
very high energy strings  occasionally
emit the zero modes (massless point particles) due to the
cosmological expansion and sometimes absorb the zero modes
due to the cosmological contraction.
This process is shown to be  adiabatic.
This adiabaticity, or in other words the reversibility,
ensures that the oscillation is not damping.
Thus our result is quite consistent with the picture of
the initially
oscillating universe
which is followed by the three dimensionally expanding epoch.

After some three dimensional directions start
 to expand with thre remaining six
dimensions being kept Plankian,
the energy of zero modes and the strings having no
winding along the three dimension (which we call
non-winding strings) grows roughly in proportion to
the expanding volume.
The temperature is shown to be fixed to be the Hagedorn temperature
in this period.
This inflation like energy growth is
possible because the highest energy strings (which we call
 winding strings) continue to supply the energy by their decay.

This epoch ends when the high energy strings decay away.
At this stage there left are the dominant
zero modes along with a few non-winding strings.
{}From that time the redshift of the zero modes get effective resulting
 in the falling down of the temperature.   The remaining non-winding
string modes are shown to be quickly decay away because of their high
specific heat.
In this way the string universe is shown to transit to the conventional
radiation dominant universe.
The exposition of this thermal history is the  main result of this paper.


Our plan of the discussion is as follows.
In the next section we explain on what setting we
proceed the discussion in this paper and  summarize
 the approximations and the
assumptions to use in
this paper.
In Sec. \ref{sec:scenario}
  we will give a brief review of the Brandenberger and Vafa's
scenario to fix the notation.
In order to follow the thermal history of the string universe,
 we need to calculate the multi
 string state density.   In the
ideal string approximation  the multi state density is evaluated
 by the single state density.  In Sec.\ref{sec:singlestate} we
 give a detailed
 discussion on the single string state density.
Especially we explain the change occurring in
 the single state density induced by the
cosmological expansion, which is of importance in
discussing the thermal history.
In Sec.\ref{sec:big} we give a remark on the value of the total
energy of our microcanonical ensemble.  This value turns out to be the
key parameter which determines the thermal history.
In Sec.\ref{sec:framework} we will provide a  framework of the
microcanonical formalism on which we can
follow the thermal history.  From  Sec.\ref{sec:non}
to Sec.\ref{sec:zero} we
will evaluate the multi state density from the single state
densities.  In these calculation a novel technique
is introduced and used extensively.
In Sec.\ref{sec:thermal} the thermal
 history of our string universe is
deduced by gathering all the knowledges
obtained in the preceding sections.
The last section is devoted to some discussions.  In the Appendendix
we will ascertain the validity of the Maxwel-Boltzmann approximation
which will be use in this paper.

\section{Setting}
\label{sec:set}

In this section we are going to make sure what kind of tools and
approximations we use in this paper.
As is well known the thermodynamical treatment needs special care for
the string theory because of
the exponentially growing state density \cite{Hagedorn,GSW}.
 One method to treat such system is to
extend the temperature to complex number\cite{Hagedorn,Frau,Sund}
 in the canonical formalism, and
 the other is to quit to use the
canonical formalism and to use the microcanonical formalism
 \cite{BV,MT,AT,Deo,BG,DeoII}.

Both methods are actually connected through the
 Laplace transformation. We
will take the latter treatment in this paper.  Our
 interest in this paper is
the fundamental string theory not a cosmic
 string.
However so as to clarify our setting it is useful to
review what is known  in the studies   of the cosmic string.

The ensemble of the cosmic or fundamental strings is in general subject to
both  the statistical mechanics and the dynamics of the theory. In the case of
the cosmic string theory the dynamics is shown to prevail over the statistics
 \cite{MT,AT}.
This is a
consequence of the following settings. First of all for the cosmic string
theory one assumes the Einstein
gravity with Robertson-Walker  metric as a background since the
relevant energy scale is not so close to the Plankian scale. One describes
 the
string as the Nambu-Goto string in the radiation or matter dominated back
ground.  Then the description of the system
 simplifies well thanks to the
one scale principle \cite{MT,AT}.
 This principle  ensures that sooner or
later the system will be attracted to a scaling solution irrespective
of the initial configuration of the strings.
 It is shown that this
behavior of the string ensemble is far from the thermal equilibrium.

However in our case of the fundamental string theory,
 it is no longer a natural
assumption that simple Einstein gravity is applicable because
 the relevant energy scale
is as high as the Plank mass.  The dynamics of the fundamental strings  are
poorly  understood at present, thus
 we  cannot proceed further as in the case  of
cosmic string.  That is why we focus on the thermodynamical analysis
 and follow the
thermal history using it in this paper.
 This strategy is essentially the one
proposed by Brandenberger and Vafa \cite{BV}.

Below we clarify our approximations  used in this paper.
We are going  to investigate actually the properties
 of the string ideal gas in
this paper.  Thus we need   some assumptions  to
identify our system
with the real universe.  We summarize them here.

First we have to assume  the validity of the
 local thermal equilibrium and
that  the ideal gas approximation are reasonably
good for the string universe in the period of interest.
If  one of  these  is not a good one,
our system of the string ideal gas is not guaranteed to be a  good
approximation of our universe.

Next we have to assume that some special roles of (quantum) gravity
 if any
 are  not
important in considering the thermal history of our universe. Of course
string theory by birth is  the quantum theory
 unifying gravity.    Therefore, some special
effects may well exist  concerning gravity.
But our present knowledge is so
poor that we have to assume their unimportance.

Even under  this difficult circumstance we do
think that it is much more meaningful to do something rather
than doing nothing.
Some foothold may well be found by such trial.
Our main purpose of this paper is to provide the
 zeroth approximation of the
whole story of the string universe.

Within these approximations the thermodynamical functions have been
calculated by some authors in several models of SST
\cite{Hagedorn,Frau,Sund,MT,Deo,BG,BV}.
In following the thermal history we actually need another assumption
to determine the history uniquely.  As the last assumption
we require that the usual mechanism
of the redshift works for the massless particles even in the Plank time.
We call this assumption a normal energy loss.  In the initial
epoch this condition is shown to be equivalent to
the equi-entropy condition which is also adopted by \cite{BV}.

\section{Cosmological scenario}
\label{sec:scenario}

In order to fix the notations we present a brief review of
 the Brandenberger and Vafa's \cite{BV,TV} scenario in this
section.

They started with the Heterotic string theory \cite{Gross} in
the nine dimensional torus $ T^9\times R$.
For this model the single string spectrum reads \cite{GSW}
\begin{eqnarray}
\varepsilon^2&=&2 r^2+4(n_R+n_L) M_s^2 \nonumber\\
r^2&=&\sum_{i=1}^9
    \left[  \left( {n_i\over a_i}\right)^2+
               \left( {m_i  a_i M_s^2}\right)^2 \right ] \nonumber\\
\label{eq:spec}
\end{eqnarray}
where
\begin{eqnarray}
 m_i=n_i&=0,\pm 1,\pm 2,\cdots \nonumber\\
  n_R,n_L&=0,1,2,\cdots\nonumber\\
 M_s&=1/\sqrt{2 \alpha^\prime}.\nonumber\\
\nonumber\end{eqnarray}
In these expressions
$\sqrt{2}\pi a_i$ is  a linear size of the torus. In the
string theory, $M_s$ which is of the order of the Plank mass
is  the only dimensionful constant.   We frequently
 set $M_s$ to unity in the sequel.

The significant feature of this model is a
duality $a \leftrightarrow 1/a$ which is manifest in the spectrum.
This symmetry connects
the large volume world with the small volume world.  This is called  the
target space duality \cite{KY}.
The self dual point of this duality is
$a=1$ (in units of $1/M_s$).  It
 is known that in the low energy limit of the closed string theory
there emerges Einstein gravity \cite{GSW}. However
 Einstein gravity does not respect this duality \cite{BV}.  This means
that the use of Einstein gravity is not legitimate in this realm.

 They considered
that Einstein gravity is modified in this
 realm so as to respect this duality.
The winding mode is thought to play an essential role in this realm.  Their
approximate estimation showed that the winding mode works so as to slow down
the expansion of the universe while the momentum mode  to slow down the
contraction. These effects
 make the universe oscillating around the self
dual point for a while \cite{TV}.

Their description of how three dimensional universe is born is as follows.
Suppose that a $d$ dimensional space out of nine gets
larger than the average by accident.   Then the winding modes along these
$d$ directions get more massive than the rest
so that they tend to decay more than
the average.  Consequently, the
 number of these modes would be reduced.
Since  the winding modes  slow down the expansion,
these $d$ directions  becomes easier to expand
than the other
 directions.  Thus
the perturbation considered above has an
unstable nature.  They suggested  that
 in this way the $d$ dimensional space
becomes much larger in size
than the Plank length while the remaining $9-d$ dimensions are kept
Plankian scale.  Since the
 Plankian scale  is invisible at low energies,
this  mechanism effectively reduces the
dimensionality of  space time; a $d+1<10$
dimensional large universe arises out of
small $T^9\times R$.  They called this mechanism a  decompactification.


\section{Single state density}
\label{sec:singlestate}

In this section we are going to investigate
  fundamental properties of the  single string state density
 $f(\varepsilon)$.
  This provides the theoretical foundation of our
discussion of  the thermal history of
the string universe.  In fact as we see in the later sections,
the thermal history is deduced from the functional form of the multi
state density and the multi state density is calculated from the single
state density.
In the high energy range the functional form of $f(\varepsilon)$ has already
 been  estimated analytically.
We first review this result and explain how its volume dependence
comes out.   Later this volume
 dependence will prove to have
 key importance in observing  the thermal history of the string universe.
Next we will give our numerical  estimation of $f(\varepsilon)$ by the direct
counting of the single string states in the low energy range.
This clarifies the explicit number distribution of the strings.
Lastly we remark that an interesting effect in the
single state density is induced as the cosmological expansion proceeds.

\subsection{High energy behavior}
\label{subsec:high}

For the general closed super string theory in a compact space which is
multiply connected, the single state density is written as
\cite{Hagedorn,Frau,Sund,MT,AT,Deo,BG,BV}
\begin{equation}
f(\varepsilon)={CV\over\varepsilon^{\eta+1} } \hbox{e}^{{\beta_H \varepsilon}}
\label{eq:single}
\end{equation}
for large enough energy $\varepsilon$.
In this expression $\eta=D/2$ with $D$ being the number of noncompact
dimensions, $V$ denotes a $D$-dimensional volume and $1/{\beta_H }  $  is a
constant called the Hagedorn
 temperature. The constants $C$ and  ${\beta_H }  $
 depend on the
string model.
The above form of the single state density is uniquely implied by the single
string spectrum (\ref{eq:spec}).
{}From (\ref{eq:spec})\ we learn
 that the energy of the string consists of a kinetic part
$r$ and an oscillation part $n_R+n_L$.
For the kinetic part $r^2$ we have winding modes $a_i m_i$ in
addition to the usual
momentum modes because it is
the closed string theory in the non-simply connected manifold.

Before considering
the volume dependence of (\ref{eq:single}), let us
 consider what form of the single state
  density is implied in the usual relativistic
 particle. Such system
 does not have the oscillation mode
and the winding mode, the spectrum (\ref{eq:spec})\  reduces to a simpler form
$\varepsilon^2=\sum_{i=1}^d \left( { n_i / a_i} \right)^2 $
on the $d$-dimensional torus.
The single state density in this case is  proportional to the surface
area of the elliptic sphere in $d$-dimension having
axes $\varepsilon a_1,\cdots, \varepsilon a_d$.  Namely we get
\begin{equation}
f(\varepsilon)\propto {d\over d\varepsilon}
\left(\prod a_i\right) \varepsilon^d=V \varepsilon^{d-1}
\end{equation}
where $V$ is a $d$-dimensional volume.

This represents a simple fact that the single state
 density is an extensive
quantity.
One of the peculiar phenomena in the string thermodynamics is that $f$ is
no longer an extensive quantity.  In
 fact the number $D$ concerning in (\ref{eq:single})
is not the total space dimension but the noncompact dimension.  This
 means that $f$ is not extensive.  In the extreme case of a totally
 compact space,
 $f$ is volume independent.

Let us see how this peculiar behavior comes out.
Only the kinetic part reflects the structure of the space, so that
 we concentrate ourselves on the degeneracy of the kinetic part. Just
like the case of the usual point particles, the
 degeneracy is obtained as
the surface area of the elliptic sphere having the axes
$a_1 r,\cdots,a_9r,r/a_1,\cdots,r/a_9$, see (\ref{eq:spec}).
The state density is therefore proportional to
\begin{equation}
 {d\over d\varepsilon}\left(a_1r\times a_2r\times
  \cdots\times a_9r\times {r\over a_1}\times{r\over a_2}\times
  \cdots \times{r\over a_9}\right).
\end{equation}
This shows that  $f(\varepsilon)$ is
 certainly $a_i$ independent. This property is
essentially a consequence of the cancellation between the momentum mode and
the corresponding winding mode.
As the volume expands the phase space of
the former increases  while
that of the the latter decreases.
Now let us see what happens if $D$ dimensions are open. This time $D$
momentum modes miss their partner to cancel, so that $a_i$ dependence
remains.
  Accordingly we get $f(\varepsilon) \propto a_1\ldots a_D$. This
 surely explains the
 peculiar volume dependence shown in  (\ref{eq:single}).

\subsection{Numerical analysis in low energy range}
\label{subsec:num}

We present here the result of our numerical analysis.  In Fig.\ref{fig1}
presented is the plot of
$f(\varepsilon)\hbox{e}^{-{\beta_H \varepsilon}}$ in the totally compact
case $D=0$.
In the current situation the kinetic energy is
discrete by the finiteness of the
space, which makes the energy spectrum discrete as seen in Fig.\ref{fig1}.
We also plot $\varepsilon f(\varepsilon)\hbox{e}^{-{\beta_H \varepsilon}}$
in Fig.\ref{fig2}.
The asymptotic behavior
$ f(\varepsilon)\hbox{e}^{-{\beta_H \varepsilon}}\rightarrow 1/\varepsilon$
is clearly seen in Fig.2.
This is the first time that this aysmptotic behavior is shown to set in
already at $\sim 10 M_s$.

These two figures in fact have a clear physical meaning.
It is shown \cite{Deo} in the microcanonical formalism that
$f(\varepsilon)\hbox{e}^{-{\beta_H \varepsilon}}$ and
$\varepsilon f(\varepsilon)\hbox{e}^{-{\beta_H \varepsilon}}$ represent
 the number distribution and
 the energy distribution of the
strings respectively.


As we will recognize later, $D=3$ case is relevant to our discussion.
The value of $f(\varepsilon)\hbox{e}^{-{\beta_H \varepsilon}}/V$ versus
$\varepsilon$ is presented in Fig.\ref{fig3}.
The spiky behavior therein represents the opening of various modes.
The analytic estimation indicates  that the quantity
tends to behave $C/\varepsilon^{5/2}$ for large $\varepsilon$
(see (\ref{eq:single})).
The plot
$\varepsilon^{5/2} f(\varepsilon)\hbox{e}^{-{\beta_H \varepsilon}}/V$
 is shown in Fig.\ref{fig4}
which
justifies this asymptotic behavior and tells
 us where this behavior  sets in.


\subsection{Change in the single state density}
\label{subsec:change}

In this subsection we examine  what
occurs to $f(\varepsilon)$ when  accidentally chosen three
directions are expanding while the remaining dimensions are
kept Plankian.  Because
we will restrict ourselves to the case in which three directions are
expanding at an equal rate, we set $a_1=a_2=a_3=a $
and $a_4=\cdots=a_9=b
\sim 1$ (see (\ref{eq:spec})) from now on.

In the preceding section we saw that the high energy behavior of the single
state density $f(\varepsilon)$ is independent of $a $ since $D=0$.  This
 is the  consequence of the
cancellation between the momentum mode and the winding mode.  However this
cancellation becomes incomplete at low energies for the following reason.

As  $a$ gets larger the winding mode along the $a$-direction is getting
heavier.    Eventually the  winding mode
 in that direction becomes too heavy to
be excited especially in the low energy range.
  This means  the
winding modes along the expanding three directions are
effectively frozen.   Then the cancellation
between the momentum mode and the winding mode breaks down in the low
energy region.    As a result the single state density
behaves as if $D=3$ instead of
$D=0$ at low energies.    Namely
$f(\varepsilon)\hbox{e}^{-{\beta_H \varepsilon}}$ behaves as
$CV/ \varepsilon^{5/2}$ at low energies
and as  $ 1/\varepsilon$ in the high energy range, respectively.
We denote this energy $m_0$ which  separates the low and high energy ranges.
As $a$ becomes large, the strings with higher energies
 behave as if $D=3$.
Namely the effective $D=3$ range extends as the universe expands.

In fact it can be  shown by examining
the functional form of the state density
that as $a$ grows $m_0$ grows at the rate
$m_0(a)\propto a^2$.
This is justified by the numerical analysis of Allega et al.'s and ours.
This phenomena has been discussed
 also by other authors, in different
ways by P.Slomonson et al in \cite{Sund} and  by authors in
 \cite{Allega,DeoII}.

Before closing this section  we summarize the behavior of
 the single state density $f$ in an expanding
epoch.
Below the first excitation energy $\varepsilon<m_1$,
$f$ describes the zero modes which is
regarded
 as the usual point particles. In the range between $m_1$ and $m_0$ the
state density behaves as if the string gas in  open three space
 dimensions. We call
the string in this range a non-winding string because such string
does not have  winding along the expanding directions.
Note that the non-winding string in general has a winding along the
remaining six directions.

Lastly for the energy greater
than   $m_0(a)$, $f$ behaves as the string
gas in the totally compact space.
We call such string as a winding string.  We
 can express the single state density
using the step function $\theta$ as:
\begin{eqnarray}
f(\varepsilon)=f_z(\varepsilon)&+&
{CV\over \varepsilon^{\eta+1}}
\hbox{e}^{{\beta_H \varepsilon}}
\theta(\varepsilon-m_1)\theta(m_0(a)-\varepsilon)\nonumber\\
&+&{1\over \varepsilon}
\hbox{e}^{{\beta_H \varepsilon}}\theta(\varepsilon-m_0(a))\nonumber\\
\end{eqnarray}
 with $\eta=3/2$, where $f_z$ denotes the state density of the zero
modes.
Although our  interest in this paper
 is in the $\eta=3/2$ case only, we keep $\eta$ arbitrary in
the subsequent expressions for better  understanding of the structure of
our treatment.
We again stress here that $f$ is proportional to $V$ below $m_0(a)$.

\section{How big should the total energy be?}
\label{sec:big}

In this section
we will give an important remark on the  question
how big the total energy $E$
of our microcanonical ensemble should be.

Let us consider the string distribution in the initial epoch.
The number distribution of
 strings are displayed in
Fig.\ref{fig1}, see the second paragraph of \ref{subsec:num}.
The first excited
 state opens at $m_1=\sqrt{8}(\times M_s)$
corresponding to $N=n_R+n_L=2$
instead of $N=1$ since the latter is inconsistent with
the level matching condition of the Heterotic string theory \cite{GSW}.

The quantity $E$
is the total energy of the microcanonical ensemble which we have to
introduce
at the beginning of the discussion.  What we can
 find from Fig.\ref{fig1} is that if
we take $E$ of the order of $M_s$ as in the
usual dimensional analysis, we
 have no string modes from the beginning since
the distribution terminates at $\varepsilon=E$.
In such situation our universe is no longer a model of the string
cosmology.

You may say that we only have to take $E$ as large as we like.   However
 in the
cosmology with causality we can only have finite region in the thermal
equilibrium
because the speed of light is finite.  We are dealing with
the equilibrium thermodynamics in this paper.  Therefore
it is implicitly assumed that the spatial
region having the energy $E$ must be in the thermal equlibrium.
Consequently we cannot take $E$ as large as we like.

  Now we define $E$ to
be the maximally allowed energy in the thermal equilibrium and
discuss how big $E$ can be.

  Since our present knowledge of the string theory does not allows us
to determine  the value of $E$, we are left with
two possibilities.
The first one is that
 the region having $E$ (defined as above)
 is smaller than the whole universe,
the second one  is that the whole torus universe is in the thermal
 equilibrium.

First we consider the former case.
Because the value of $E$ is considered to be the maximal
 energy which a
single string can occupy,
it  appears that $E$ must be large enough in
order for our universe to be regarded as a model of the string cosmology.
But it is not necessarily the case.  In fact there is
 a loophole in this argument.
A string extending over beyond the causal region
can have the energy  much greater than $E$.
Such acausal fundamental
string may well be produced if the universe itself is born through a quantum
tunneling or something like that.  However
 such situation is beyond  control of our present technology. We do
not and cannnot go further in such case in this paper.

Next we consider the second case when the
 whole universe is in the thermal equilibrium.
Now we simply conclude that the value of $E$ must be large for the
universe to be full of strings.  There is no loophole this time.

Because the loophole  in the former possibility is out of
our control we decide to assume that
$E$ is large enough in this paper.
Of course the alternative case that our universe has no strings
even in the initial epoch is
another possibility.  However we will not
 treat this case since the purpose of this
paper is to explore the possibility of the  universe full of strings.

As we stated before our scenario is based on  Brandenberger and
Vafa's  one in
which the string
universe is supposed to oscillate  around the self dual point for
the initial period.  This picture nicely
 fits to the second case mentioned above, since
the oscillation over many periods
tends to thermalize the whole
universe.   Even if we had started in the loophole case, this
 oscillation makes the acausal strings causal.
Moreover, we point out that this picture has another advantage from the
cosmological view point.  This picture
is plausible from the view point of the
horizon problem. If the whole universe would be thermalized during
 the initial
period of the oscillation,  we would go  through the history
 with the background radiation
of the same temperature at any part of the universe.

\section{Framework to follow the thermal history}
\label{sec:framework}

In this section we  intend to give a
concrete framework to follow the thermal
history of the string universe. Many authors have discussed the behavior
of the string gas in the micro canonical formalism so
far\cite{Hagedorn,Frau,Sund,MT,AT,Deo,BG,BV}. The
str has been
 examined in various ways and the differences lying
between the  stringy phase (which
 is frequently referred to as a high density
phase) and the low temperature phase
 (which is referred to as a low density phase)
have been exposed.

However very little of
the attention have been paied to the transition of
these two phases. The problem how the stringy universe evolves
into  the
radiation dominated universe is still an open problem.
In order to treat the  transient period we present  a concrete
 framework based on  the microcanonical formalism.

The multi state density  is written by the
single state density under the Maxwell-Boltzmann (M-B) approximation
as \cite{Deo}
\begin{equation}
\Omega(E)=
  \sum_{n=1}^\infty {1\over n!}\int_0^\infty
       \prod_{j=1}^n d\varepsilon_j f(\varepsilon_j)
   \delta\left(E-\sum_{j=1}^n \varepsilon_j\right) .
\label{eq:multi}
\end{equation}
We discuss the validity of the M-B approximation in the Appendix.
\widetext
We recall that the single state density $f$ is expressed as a sum of the state
densities of the zero modes, the non-winding strings and the winding
strings. We define the multi state densities associated
 with  these regions by imitating
(\ref{eq:multi}) as:
 \begin{eqnarray}
\omega_z(\varepsilon_z)&=&
\sum_{n=1}^\infty {1\over n!}\int_0^\infty
       \prod_{j=1}^n d\varepsilon_j f_z(\varepsilon_j)
              \delta \left(\varepsilon_z-\sum_{j=1}^n \varepsilon_j\right),
                   \nonumber\\
\omega_N(\varepsilon_N)&=&
       \sum_{n=1}^\infty {1\over n!}\int_0^\infty
       \prod_{j=1}^n
       {CV d\varepsilon_j \over \varepsilon_j^{\eta+1} }
\hbox{e}^{{\beta_H }\varepsilon_j}
\theta(\varepsilon_j-m_1)
 \theta\left( m_0(a)-\varepsilon_j \right)
    \delta\left(\varepsilon_N-\sum_{j=1}^n
 \varepsilon_j\right), \nonumber\\
\omega_W(\varepsilon_W)&=
&\sum_{n=1}^\infty {1\over n!}\int_0^\infty
       \prod_{j=1}^n
       { d\varepsilon_j \over \varepsilon_j }e^{{\beta_H }\varepsilon_j}
\theta\left(\varepsilon_j-m_0(a)\right)
 \delta\left(\varepsilon_W-\sum_{j=1}^n \varepsilon_j\right).
 \nonumber\\
\label{eq:omegas}
\end{eqnarray}

\narrowtext
The similarity between (\ref{eq:multi}) and the
 Taylor
expansion of the exponential function enables us to anticipate that
$\Omega$ is expressed as a product form of $\omega$'s.
Actually we can prove by a straightforward calculation that
\begin{eqnarray}
\hat\Omega(E)=
\int_0^\infty d{\varepsilon_z} &d&{\varepsilon_N} d{\varepsilon_W} \,\,
  \hat\omega_z({\varepsilon_z})
\hat\omega_N({\varepsilon_N})
\hat\omega_W({\varepsilon_W})\nonumber\\
&\times&\delta(E-{\varepsilon_z}-{\varepsilon_N}-{\varepsilon_W}).\nonumber\\
\label{eq:hmulti}
\end{eqnarray}
The carets on the top of $\omega$'s  mean the
 addition of the delta function,
$\hat\omega(\varepsilon)=\omega(\varepsilon)+\delta(\varepsilon)$.
These carets
 enable us
 to express the equation in a simple form as above. The necessity of
the delta functions is readily
 understood if we recall that right hand side of
(\ref{eq:hmulti}) counts
the number of all the composite
 states of three kinds of substances the zero modes,
the non-winding strings and the  winding strings.
For example there are also the states having no zero
modes,  which must be counted.
  The term $\delta({\varepsilon_z})
\omega_N({\varepsilon_N})\omega_W({\varepsilon_W})$  is
 responsible for these states to be taken into account in the integration.

Based on the equation we argue as follows. Because the delta function ensuring
the energy conservation is included in the right hand
side of (\ref{eq:hmulti}), we can perform an
integral over $ {\varepsilon_W}$ to obtain
 a two dimensional integration over $({\varepsilon_z},{\varepsilon_N})$.
In many cases of  interest
 the integrand  has a sharp peak at some single point on
the $({\varepsilon_z},{\varepsilon_N})$ plane, and the contribution from
this point dominates the integral.
The position of the peak is dependent on $a$ since the $\omega$'s have an
implicit  dependence on $a$. We denote the
 position of the peak as $(e_z(a),e_N(a))$.

If we recall the fundamental principle of the equal a priori probability, we
conclude that we find the subsystem in the energies $(e_z(a),e_N(a))$ when
the size of the universe is $a$.
This is because this state has
 an overwhelming probability.
 This is nothing but the essence of the microcanonical formalism.

Therefore, once we find the functional form of $e_z(a)$ and $e_N(a)$,
we can
follow the thermal history of the universe. This is our
strategy to determine the
thermal history in the microcanonical formalism.
To carry out this program
 we need to calculate the multi string state densities.
In the next three sections we will evaluate the
multi state densities associated with non-winding strings, winding
 strings and the zero modes  successively.

\section{Multi state density of the non-winding strings }
\label{sec:non}

The evaluation of the multi state densities
$\omega_N(\varepsilon)$ and $\omega_W(\varepsilon)$ have been given  by
several authors \cite{MT,Deo,BG,BV}.
The method used there is the Laplace transformation and the
saddle point approximation.  We note that  the latter is reliable only for
 large $\varepsilon$.
However as we will see later that the small $\varepsilon$
behavior of $\omega_N(\varepsilon)$ is
 necessary in the analysis of the thermal
history.

In the following we will employ a completely new method
 to examine the form
of $\omega_N$ and $\omega_W$.
That is the characterization  of $\omega$ 's by a
differential-difference equation.
We will show that $\omega$'s are solutions to some linear
differential-difference equations, and
 solve them to find the functional form of
$\omega$'s.

\subsection{Evaluation}
\label{subsec:nonmulti}

\widetext
First we remark that we can factor out the exponential part
of $\omega_N$ as
  $\omega_N(\varepsilon)=A(\varepsilon,v)
\hbox{e}^{{\beta_H \varepsilon}}$ with
\begin{equation}
A(\varepsilon,v)=
\sum_{n=1}^\infty {1\over n!}\int_0^\infty
       \prod_{j=1}^n
       {v d\varepsilon_j \over \varepsilon_j^{\eta+1} }
\theta(\varepsilon_j-m_1)
\theta\left(m_0(a)-\varepsilon_j\right)
\delta\left(\varepsilon-\sum_{j=1}^n \varepsilon_j\right).
\label{eq:omegabar}
\end{equation}
where  $v=CV$.
This is possible since the integration is constrained by the delta function.
If  we denote
\begin{eqnarray}
A_n(\varepsilon,v)=
 \int_0^\infty
       \prod_{j=1}^n
       { d\varepsilon_j \over \varepsilon_j^{\eta+1} }
  &\theta&(\varepsilon_j-m_1)
 \theta\left(m_0(a)-\varepsilon_j\right)\nonumber\\
 \times &\delta&\left(\varepsilon-\sum_{j=1}^n \varepsilon_j\right),
   \nonumber\\
\label{eq:bndef}
\end{eqnarray}
 (\ref{eq:omegabar}) is rewritten as
\begin{equation}
  A(\varepsilon,v)=\sum_{n=1}^\infty {v^n\over n!}A_n(\varepsilon,v)
\end{equation}
We change the variables as
 $\varepsilon_j\rightarrow \varepsilon x_j$ and obtain
\begin{equation}
A_n(\varepsilon,v)={1\over \varepsilon^{\eta n+1}} \int_0^\infty
       \prod_{j=1}^n
       { dx_j \over x_j^{\eta+1} }
  \theta(x_j-m_1/\varepsilon)
 \delta\left(m_0(a)/\varepsilon-x_j\right)
  \theta\left(1-\sum_{j=1}^n x_j\right).
\end{equation}
By operating $\varepsilon\partial_\varepsilon$ and
 $\eta v\partial_v$ on it we have
\begin{eqnarray}
\varepsilon&\partial_\varepsilon& A_n(\varepsilon,v)
=-(\eta n+1)  A_n(\varepsilon,v)+
    {n\over m_1^\eta} A_{n-1}(\varepsilon-m_1,v)-
    {n\over m_0^\eta} A_{n-1}(\varepsilon-m_0,v),\nonumber\\
\hbox{and}\,\,\, &    &  \nonumber\\
\eta v &\partial_v& A_n(\varepsilon,v)
={n\over m_0^\eta} A_{n-1}(\varepsilon-m_0,v),
         \nonumber\\
\end{eqnarray}
respectively. Here we made use of the fact that
$m_0(a)\propto a^2$ $ \propto v^{1/\eta}$ (see Sec.
\ref{subsec:change}).

\narrowtext
{}From these we obtain
\begin{eqnarray}
(1+&\varepsilon&\partial_\varepsilon +\eta v\partial_v )
  A_n(\varepsilon,v)\nonumber\\
&=&-\eta n A_n(\varepsilon,v)+
    {n\over m_1^\eta} A_{n-1}(\varepsilon-m_1,v) .\nonumber\\
\label{eq:rec}
\end{eqnarray}
Summing up this equation over all $n$ we get the equation
\begin{equation}
(1+\varepsilon\partial_\varepsilon +\eta v\partial_v )
 A(\varepsilon,v)=
    {v \over m_1^\eta} A(\varepsilon-m_1,v)  .
\label{eq:ddeq}
\end{equation}
 This is the equation exactly satisfied by $A$.

When $\varepsilon $ is large compared with $m_1$, we can use an approximation
$A(\varepsilon-m_1,v)$
$=A(\varepsilon,v)-m_1\partial_\varepsilon A(\varepsilon,v)$
 to rewrite the equation
as
\begin{equation}
(1+ \varepsilon\partial_\varepsilon+ {v\over m_1^\eta}m_1\partial_\varepsilon
   +\eta v\partial_v)A(\varepsilon,v)={v\over m_1^\eta}A(\varepsilon,v).
\label{eq:aps}
\end{equation}

In the case that string has high energy
density $\varepsilon/m_1>>{v \over\eta m_1^\eta}$
( which is the case
studied in refs.\cite{MT,Deo,BG}, we can neglect the third term of the
left hand side of (\ref{eq:aps}) in comparison with the second term.
The equation reduces to
\begin{equation}
( 1+\varepsilon\partial_\varepsilon
   +\eta v\partial_v)A(\varepsilon,v)
={v\over m_1^\eta}A(\varepsilon,v).
\end{equation}
This can be solved with ease to obtain
\begin{eqnarray}
A(\varepsilon,v)&=&
{v\over \varepsilon^{\eta+1} }
g^\prime\left(-{v\over \eta \varepsilon^\eta} \right)\nonumber\\
&\times&\hbox{exp}\left[  g\left(-{v\over \eta \varepsilon^\eta} \right)+
{v\over \eta
m_1^\eta}  \right]\theta(\varepsilon-m_1),\nonumber\\
\end{eqnarray}
with some analytic function $g(x)$.
The last step function
 means that $A(\varepsilon,v)=0$ for $\varepsilon\leq
 m_1$ by
definition.
Comparing this with  the $\eta=0$ solution which
will be obtained in the  next section, we can
 determine $g$ up to $x$ as
\begin{equation}
g(x)=0+(\eta O(\eta)+1)x+ \cdots,
\end{equation}
where $O(\eta)$ represents the function of $\eta$ vanishing at $\eta=0$.

Consequently for the string gas which has the
 energy ${\varepsilon_N}$
 such that $v/\eta \varepsilon_N^\eta\sim 0$,  the multi state density is
expressed as
\begin{equation}
\omega_N(\varepsilon,v)=const\times {CV \over \varepsilon^{\eta+1}}
\hbox{exp}\left(  {CV\over \eta m_1^\eta}+{\beta_H \varepsilon}  \right).
\end{equation}
This reproduces the known result \cite{MT,Deo,BG}.

We have examined the high density region above.
However, we will realize that for our string universe
the region with low energy density
$\varepsilon/{m_1}<<{v/\eta m_1^\eta} $ is relevant.  Thus
we have to evaluate $\omega_N(\varepsilon,v)$ in this low energy density.
Before carrying it out we estimate the position of the peak
of $A$ and the height of it which are of importance in the
thermal history.

Setting $\varepsilon=e_z(a)$ in (\ref{eq:aps})  this equation is reduced
to an
ordinary differential equation
\begin{equation}
(1+\eta v\partial_v)A(e_N(a),v)= {v \over\eta m_1^\eta} A(e_N(a),v),
\end{equation}
since $\partial_\varepsilon A$ vanishes on that point.
This ordinary differential equation is readily solved and we
get the height of the peak:
\begin{equation}
A(e_N(a),v)
={const\over
 v^{1/\eta}}\hbox{exp}\left[{v\over \eta {m_1}^\eta}\right].
\label{eq:entn}
\end{equation}

The logarithm of this is nothing but the entropy of the non-winding strings.
Therefore,  the expression simply tells us that the entropy produced
when $D$ dimensional volume out
of nine  expands is proportional to the expanding
volume (note that $v=CV$).

Next we  determine the functional form of $e_N(a)$.
{}From a numerical determination of $A(\varepsilon,v) $ which we will
give later, we see
 that
$v/\eta {m_1}^\eta $ is much greater than $e_z(a)/{m_1}$.
In  this case we can neglect
 the second term instead of the third one in (\ref{eq:aps}).
This reduces the equation to
\begin{equation}
\left(1+{v \over {m_1}^\eta}{m_1} \partial_\varepsilon+\eta
v\partial_v\right) A(\varepsilon,v)
     ={v\over {m_1}^\eta}A(\varepsilon,v).
\end{equation}
By solving it we can determine the functional form of $A(\varepsilon,v)$
in this region as
\begin{equation}
A(\varepsilon,v)={1\over v^{1/ \eta}}
  \hbox{exp}\left[h\left({\varepsilon\over {m_1}}-{v \over\eta m_1^\eta}
\right)+{v \over\eta m_1^\eta} \right],
\label{eq:dimadv}
\end{equation}
where $h$ is some function to be determined by a boundary condition.
But the explicit form of $h$ is unnecessary for our present purpose.
 The function $A(\varepsilon,v)$ is maximized  when the function
$h(x)$ is maximal.  Denoting the position of the
peak of $h(x)$  as $x=c$, we can express the position of the peak
of $A$, $e_N(a)$ as
\begin{equation}
e_N(a)/{m_1}={v \over\eta m_1^\eta}+c.
\label{eq:upfour}
\end{equation}
Numerically it can be expressed as  $e_N(a)=4.00 a^3+const$.
Namely $e_N(a)$ increases in proportion to the volume for large $a$.

We solved (\ref{eq:ddeq}) for small $\varepsilon$, at growing values of $a$.
Fig.\ref{fig5} shows the
 plot of $\ln A(\varepsilon,v)$ versus $\varepsilon$.
  As the energy increases, new modes  start to
open.
This fact is exhibited in the
low energy behavior of $A(\varepsilon,v)$ as the
emergence of
several peaks.  We see that
 the position of the peak moves to higher energy as $a$ grows.
In high energy region, the spiky behavior seen at
  low energies
is smeared, resulting in  a smooth curve.
Fig.\ref{fig6} and Fig.\ref{fig7} show the plots of $(e_N(a)/{m_1})/a^3$ and
$\ln( A({\varepsilon_N},v) )/({v/\eta m_1^\eta})$ versus $a$, respectively.
The behaviors in both figures
 are consistent with the above analytic estimate.

To prepare for the later usage we present here the plots
of a microcanonical temperature of the non-winding strings.
This  quantity defined as
\begin{equation}
  \beta_N(\varepsilon,v)=
\partial_\varepsilon\ln \omega_N(\varepsilon,v)=
\partial_\varepsilon\ln A(\varepsilon,v)+{\beta_H }
\label{eq:microM}
\end{equation}
measures the rate of the entropy increase
 due to the energy increase.
We show the plots of
 $  \beta_N/{\beta_H }$ versus $\varepsilon$ in Fig.\ref{fig8}.

The global decreasing behavior means that the specific heat is globally
positive.  The point of $\varepsilon$ on which $\beta_N={\beta_H }$ is where
$\ln A(\varepsilon, v) $ peaks.
The local spiky behavior around
$\varepsilon\sim m_1$ means the thermodynamical
instability in this region.


\section {Multi state densities of the winding strings}
\label{sec:windmulti}

Next we calculate the multi state density of the winding strings. We only
have to make an analogue of the previous discussion.
Since $\eta=0$  in the present case we get
\begin{equation}
(1+\varepsilon\partial_\varepsilon )A(\varepsilon)=A(\varepsilon-m_0)
\end{equation}
instead of (\ref{eq:ddeq}).
Using a similar approximation we have $\partial_\varepsilon A(\varepsilon)=0$.
Determining the normalization by
\begin{equation}
A(m_0)=
\int_{m_0}^\varepsilon {d\varepsilon_1\over \varepsilon_1}
 \theta(m_0-\varepsilon_1)\delta(\varepsilon-\varepsilon_1)
\big\vert_{\varepsilon=m_0}=1/m_0,
\end{equation}
 we finally obtain
\begin{equation}
\omega_W(\varepsilon)
={1\over  m_0}\hbox{e}^{\beta_H \varepsilon} \theta(\varepsilon-m_0).
\label{eq:flat}
\end{equation}
This reproduces the known result \cite{BV}.
One particular feature of this functional form is that
$\omega_W(\varepsilon)\hbox{e}^{-{\beta_H \varepsilon}}$ has no peak.  This
 is not the case for
the non-winding strings and the zero modes.


\section{Multi state density of the zero modes }
\label{sec:zero}

In this section we estimate the multi state density of the zero modes.
When we enter into the  thermal history of the string universe,
the knowledge on
the position of the peak of the function
 $\omega_z(\varepsilon,v)\hbox{e}^{-{\beta_H \varepsilon}}$
will be necessary.  We will determine this in the last part of this section.

The multi state density $\omega_z(\varepsilon,v)$ in (\ref{eq:omegas}) can be
 rewritten as
\begin{eqnarray}
\omega_z&(&\varepsilon,v)
=\hbox{e}^{\beta \varepsilon}\nonumber\\
&\times&\sum_1^\infty {1\over n!}\int_0^\infty
  \prod_1^n d\varepsilon_j
 f_z(\varepsilon_j)\hbox{e}^{-\beta \varepsilon_j}
 \delta\left(\varepsilon-\sum_{j=1}^k \varepsilon_j\right),
    \nonumber\\
\label{eq:zero}
\end{eqnarray}
with an arbitrary positive $\beta$ parameter noting
the delta function constraint.

In order to estimate the power series above, we note two facts here. First
the sum of the exponential function  $\sum {x^k / k!} $ receives its
dominant contribution from $k=x$ term.
In fact it can be verified that the expression ${x^k / k!} $ if
seen as a
function of $k$ has a sharp peak at $k=x$.

Next we can ascertain that
 the power series of $\omega_z (\varepsilon)$ can be regarded as
 $\sum x^k/k!$.   In order for it to be valid
it is sufficient that
 the important contribution to the
integral of $\varepsilon$'s  comes from the diagonal
 region $\varepsilon_1\sim\varepsilon_2\sim\cdots\sim\varepsilon_k$.
As we mentioned
before $f_z$ behaves as a positive power of $\varepsilon$, see the
Sec. \ref{subsec:high}. This
 enables us to  apply
the well-known inequality,
$\root k \of {\varepsilon_1\varepsilon_2\cdots\varepsilon_k}
\leq {1\over k}\Sigma \varepsilon_j$
to conclude that $\prod f_z(\varepsilon)$ maximizes on the diagonal region
justifying our claim.

Because $x$ of
$\sum x^k/k!$ corresponds to $x_0=\int\limits_0^{\infty} f_z
(\varepsilon) e^{-\beta\varepsilon} d\varepsilon$ of (\ref{eq:zero}),
we only have to   focus on the
$k=x_0$ term.
Let us examine for which energy $\varepsilon$  this term does not vanish.
For the
delta function not to vanish, the argument of it must vanish.
So let us see what is the typical value of $\sum_{j=1}^k \varepsilon_j$ in
the delta function of (\ref{eq:zero}).
This is estimated as
\begin{eqnarray}
<\sum_{j=1}^k \varepsilon_j > = k <\varepsilon_j>&=&k
{ \int\limits_0^{\infty}
d\varepsilon\varepsilon f_z(\varepsilon)e^{-\beta\varepsilon}     \over
\int\limits_0^{\infty} d\varepsilon
f_z(\varepsilon) e^{-\beta\varepsilon}      }\nonumber\\
&=&\int\limits_0^{\infty} d\varepsilon \varepsilon
f_z(\varepsilon)e^{-\beta\varepsilon}.\nonumber\\
\end{eqnarray}
We define here $h$ and $n$ as
\begin{eqnarray}
h(\beta,v)
&=&\int^\infty_0 \varepsilon
 f_z(\varepsilon)e^{-\beta\varepsilon} d\varepsilon ,\nonumber\\
n(\beta,v)&=&\int^\infty_0  f_z(\varepsilon)e^{-\beta\varepsilon}
 d\varepsilon \nonumber\\
\label{eq:eich}
\end{eqnarray}
 for later convenience.
As a result we realize that
$\omega_z(\varepsilon,v)e^{-\beta\varepsilon} $ has a sharp peak at
$\varepsilon=h(\beta,v)$ with the height $\hbox{exp}( n(\beta,v))$ with
 some width $q(\beta,v)$.

Making use of a function $G(x)$ having a peak at
 $x=0$ with a unit
width  and a unit height  we can express
$\omega_z(\varepsilon,v)e^{-\beta\varepsilon}$ as
\begin{equation}
\omega_z(\varepsilon,v)e^{-\beta\varepsilon}=
G\left(  { \varepsilon-h(\beta,v) \over q(\beta,v)} \right)
 \hbox{exp}\left( n(\beta,v)\right).
\label{eq:omZ}
\end{equation}
Especially in the case $\beta={\beta_H }$ will be relevant in the
later application.
We write it here with the explicit numerical coefficient
\begin{equation}
h({\beta_H },v)=65.0 a^3.
\label{eq:rokugo}
\end{equation}

It is worth noting that this argument does not apply to
the evaluation of $\omega_N(\varepsilon,v)$.  As we saw  in
(\ref{eq:single}) the single state density was a product of the exponential
part and the negative power of $\varepsilon$.  The exponential part is
irrelevant since it can be factored out as usual.
While the fact that the remaning part is a negative power of $\varepsilon$
implies the situation opposite to the above case since
$1/(\varepsilon_1\varepsilon_2\cdots\varepsilon_k)$ maximizes in the boundary
region such as $\varepsilon_1\sim \varepsilon$,
$\varepsilon_1\sim\varepsilon_2\sim\cdots\sim\varepsilon_1\sim m_1$.
The is the essence of what called Frautshi Carlitz picture.


\section{Thermal history}
\label{sec:thermal}

In this section we explicitly follow the thermal history of our system using
the multi state densities calculated in the previous sections.
The history which we are going to describe below consists of two distinct
epochs which we refer to as epoch I and epoch II, respectively.

We  insert $\omega$'s obtained above
into \ref{eq:hmulti}.
Among the constituents of $\Omega$, $\omega_N
$ and $ \omega_W$ have the same exponential dependence
$\hbox{e}^{{\beta_H \varepsilon}}$.
As for the zero modes, we can formally factor
out the same exponential
using the previous formula (\ref{eq:omZ}) with $\beta={\beta_H }$.

If we insert these $\omega_z$, $\omega_N$ and $\omega_W$ into
(\ref{eq:hmulti}), the
exponential factors can be combined to form $\hbox{exp}({\beta_H } E)$
thanks to the delta function
$\delta(\varepsilon-\varepsilon_W-\varepsilon_z-\varepsilon_N)$
  and then be picked out from the
integral as
\begin{eqnarray}
\hat \Omega (E,v)
 =&{1\over m_0}&
\hbox{exp}\left[{\beta_H } E+n({\beta_H },v)\right]
  \int_0^{\infty}
d\varepsilon_z d\varepsilon_N d\varepsilon_W\nonumber\\
&\times& G\left(
{ \varepsilon_z-h(\beta_H,v) \over q(\beta_H,v) }\right)
\hat A(\varepsilon_N,v)\nonumber\\
&\times&\left(\theta(\varepsilon_W-m_0)+\delta(\varepsilon_W)\right)\nonumber\\
& \times& \delta(E-\varepsilon_z-\varepsilon_N-\varepsilon_W).\nonumber\\
\label{eq:omegath}
\end{eqnarray}

\subsection{Oscillating epoch and epoch I}
\label{subsec:epochI}

As we mentioned before we find the system with
 energies $\varepsilon_z$, $\varepsilon_N$,
and $\varepsilon_W$ are determined as the position of the peak of the
integrand. The functions $G$ and $\hat A$
prefer that ${\varepsilon_z}$ and ${\varepsilon_N}$ take
 their most probable values,
respectively.  The winding string energy $\varepsilon_W$ is adjusted to meet
the requirement by the delta function since the integrand of
(\ref{eq:omegath}) does not have other dependence on $\varepsilon_W$.

The function for the zero modes strongly favors
\begin{equation}
\varepsilon_z=e_z(a)=h(\beta_H,v)
=\int_0^\infty d\varepsilon\varepsilon
f_z(\varepsilon)\hbox{e}^{-{\beta_H \varepsilon}},
\label{eq:zerogrow}
\end{equation}
see (\ref{eq:eich}).
While the favorable value for the non-winding
strings is similarly determined as
$ \varepsilon_N=e_N(a)=m_1 v/(\eta m_1^\eta)+const$, which is derived in
(\ref{eq:upfour}).
Accordingly the most probable  value of $\varepsilon_W$ is determined as
$\varepsilon_W=e_W(a)=E-e_z(a)-e_N(a)$.

Because both $e_z(a)$ and $e_N(a)$ grows  as
$a^3$ for large $a$,  the
 ratio of $e_N(a)$ to $e_z(a)$ approaches to a constant.  Numerically,
however,
we see that $e_N(a)<<e_z(a) $ from (\ref{eq:upfour}) and (\ref{eq:rokugo}).
Namely the zero modes is always dominant over the non-winding strings.
In view of  $e_z(a)$ in (\ref{eq:zerogrow}),  we
 can find that the zero modes are
distributed in the canonical distribution with the Hagedorn temperature
$1/{\beta_H }$.  Namely the temperature in  this period
is fixed to be the Hagedorn temperature.
Therefore, the total energy of the zero modes grows in
proportion to the volume.  The
 reason why it is possible is that the winding
 strings continuously  supply the energy by decay.
The supplied energy is also given to the non-winding strings,
 resulting in the growth of $e_N(a)$ found  at (\ref{eq:upfour}).
The  conversion of the energy into the
non-winding strings and the zero modes
 continues until the winding strings disappear
 (i.e. $e_W(a)=0$).
We call the period epoch I before their disappearence.

Here we give an important comment on the change of the total energy
$E$.   Generally in cosmology the total energy is not a constant
quantity \cite{early}.  For example the energy density of the radiation
dominated universe scales as
\begin{equation}
e_z(a)/a^3\propto 1/a^4
\end{equation}
implying $e_z(a)\propto 1/a$.   This
 energy loss is attributed to the redshift of
  the radiation due to the expansion
 of the universe.  If the size of the universe is
multiplied by a factor $a$, the wavelength is
 multiplied by the same factor.  The radiation loses its
 energy by this effect.
The  energy lost is given to the gravitational
field.   We call it a normal energy loss.

This energy loss is deduced  from Einstein gravity.  It is true that in the
period in question Einstein gravity is not a reliable approximation
because of the possible string corrections.  However
 even in this period, we consider
 that this energy loss works for the zero mode
sector.
 Then we  decide to take the
normal energy loss as the assumption based on which we follow the thermal
history of the universe.
It is amazing to observe that  this
normal energy loss  nicely fits to our
scenario.
We will show it below.

In the pure radiation case the normal energy loss is described in the
differential equation as $de_z(a)/d\varepsilon=-e_z(a)/a$.  In
 order to apply this
to our case we have to take the existence of the other modes into
account.  The energy exchange between the other modes and the
zero modes is allowed while the energy loss is only through
the redshift of the zero modes $- e_z(a)/a $.  Therefore the normal
energy loss now means
\begin{equation}
{d\over da}(e_z(a)+e_N(a)+e_W(a))={d\over da}E(a)=-{1\over a}e_z(a).
\label{eq:normal}
\end{equation}
Adding this equation to the previously given conditions
 we can uniquely determine the
functional form of $e_z(a)$, $e_N(a)$, $e_W(a)$  and $E(a)$ throughout the
thermal history.

In order to reveal what (\ref{eq:normal}) means we
examine the entropy change of the system under this assumption,
the entropy of this system is estimated from (\ref{eq:omegath}) as
\begin{equation}
S={\beta_H } E(a)+n({\beta_H },v)+\ln A(e_N(a),v).
\label{eq:totent}
\end{equation}
The last term comes from the non-winding strings.
Let us first examine the case without it.
  Then the change of $S$ due to the growth of
$a$ reads
\begin{equation}
{d\over da}S=-{\beta_H }{e_z(a)\over a}+{3 n({\beta_H },v)\over a}.
\end{equation}
Surprisingly we can show that it vanishes.   This is verified by combining
the fact that $-\partial_{\beta_H } n({\beta_H },v)$
$=h({\beta_H },v)=e_z(a)$ and
 $n({\beta_H },v)\propto 1/{\beta_H }^3$ (see (\ref{eq:eich})).
These two imply $e_z(a)={3\over {\beta_H }} n({\beta_H },v)$ which implies
$dS/da=0$.

Consequently, we have proved that as
 far as we neglect the entropy of the
 non-winding strings, the assumption of the normal energy loss is
 equivalent to
the equi-entropy.  Let us consider below what this fact means.

The generation of non-winding strings is  rephrased as the
unknotting of the winding strings along the expanding direction.
It was the
necessary condition for our universe to  exit out
 of the oscillation epoch and
entering into the three dimensionally expanding universe due
to the Brandenberger and Vafa's instability.
It is natural to suppose the universe is oscillating around the
self-dual point,
until the condition for the unknotting along some
 three directions is met.

On the other hand what we proved above is that the process
in such epoch is  adiabatic.
In other words  the process is reversible.
Sometimes  winding strings decay by generating
 the zero modes as the universe expands
 and sometimes  the winding
  strings absorb the energy from the zero modes as
the contraction proceeds.  These processes  can repeat themselves
 because they are  reversible
processes.
This situation is naturally
 identified with  the  oscillating epoch of Brandenberger and
Vafa's scenario.

Once the unknotting in some three directions proceeds enough,
the entropy generation
occurs as we have shown above.  This
 time we cannot go back because the entropy is
generated.
Namely the universe is destined to the three dimensionally
expanding universe.  This  is  natural as
 we naively expect that the birth of the three dimensional
universe  is a irreversible process.
Consequently, we have recognized that the result of our thermal analysis is
perfectly consistent with our cosmological scenario thus far.

The epoch I ends when all the winding strings decay away, in other
words when all the winding along the three directions unknot.
The point $a=a_0$ when the epoch I ends is determined by solving the
equation  $E(a)=e_z(a)+e_N(a)$.
We can easily determine the functional form of $E(a)$ from (\ref{eq:normal})
 as
\begin{equation}
E(a)=E_0+{1\over 3}(e_z(1)-e_z(a))
\label{eq:Ezero}
\end{equation}
where $E_0=E(1)$ is the initially given total energy.
Using  this, (\ref{eq:upfour}) and (\ref{eq:rokugo})   we obtain
\begin{equation}
a_0=
\left( {E_0+{1\over 3}e_z(1) \over
  {4\over 3}e_z(1)+e_N(1) } \right)^{1/3}
 \cong \left( {3 E_0 \over 4 e_z(1)}\right)^{1/3}.
\end{equation}

\subsection{Epoch II}
\label{subsec:epochII}
We call the period $a\geq a_0$ an
 epoch II.  In the epoch II there is no energy supply
 from the winding strings.   The non-winding strings and the
zero modes compete for the limited amount of total energy $E(a)$ in this time.
 The three functions
$E(a)$, $e_z(a)$ and $e_N(a)$ in this epoch is uniquely
determined by
the following three conditions:

\begin{eqnarray}
E(a)&=e_z(a)+e_N(a)\\
\beta_z(e_z(a),v)&=\beta_N(e_N(a),v)\\
{d\over da}E(a)&=-{1\over a}e_z(a) .
\end{eqnarray}

The functions in the second
 lines are the microcanonical temperature defined
as $\beta_z(\varepsilon,v)=\partial_\varepsilon \ln \omega_z(\varepsilon,v)$
 and (\ref{eq:microM}).
The second equation is  an equi-temperature condition.
As we stated before the most probable value of the energy is determined as
a meeting point of the competetion between the zero modes and the
non-winding
strings.
The functions $\beta_z$ and $\beta_N$
measure how strongly  the respective modes compete for the limited
total energy.

These equations provides the rule for the
competition between the winding strings and the zero modes.
Now the thermal history in the epoch II can be understood
qualitatively with
the knowledge of
$\beta_N$, which is given Fig.\ref{fig8}.

As  $a$ grows, the temperature of the zero modes decreases due
to the redshift.
For the equi-temperature condition to be met, the non-winding strings must
cool by the same amount.  From Fig.\ref{fig8} we can readily find that
 the non-winding strings have very large specific heat.
A very little change of the temperature corresponds to  a large  energy
change.   Thence the
decrease of the temperature signifies the violent loss of its energy.
For this reason the string modes surviving in the epoch II quickly
decay away as the  expansion proceeds.
After the strings die away
the universe come into the usual radiation dominated
universe with usual redshift
$ E(a)=e_z(a) =1/a$.

\section{Discussion}
\label{sec:discussion}

In this paper we have performed the thermodynamical analysis of the
 string cosmology focusing on
the Brandenberger and Vafa's scenario.  Our analysis was based on
the following assumptions.

\begin{itemize}
\item  The local thermal equilibrium.
\item Ideal gas approximation.
\item Normal energy loss.
\item No unexpected effect due to the non-Einstein correction.
\end{itemize}

As a result our analysis
has  presented the following thermal history of the string
universe.
In the very initial epoch
Brandenberger and Vafa's scenario suggested that the
 universe  oscillates around the self-dual
point of the target space duality.
Our analysis has shown that
the emission of the zero modes from the
strings due to the cosmological expansion and the absorption of it
due to the contraction in the initial universe are adiabatic processes.
Then these could be repeated, resulting the oscillatory behavior.
These observations clarify the nature of the   oscillating period
from the thermodynamical point of view.

Once the accidental three dimensional expansion is
  triggered, non-winding strings are produced with the entropy
production.
In this process the winding strings decay producing the non-winding
strings and the zero modes.
During this period the temperature is fixed to be the Hagedorn temperature,
realizing the inflation-like situation.

This period ends when the winding modes are exhausted.
At that time the temperature begins to fall  due to the assumed redshift
of the zero modes.
The surviving string modes  quickly die away
by this cooling process and there emerged the
usual radiation dominant universe
of the standard big bang cosmology.

Brandenberger and Vafa also addressed the problem how our space time
dimensionality is determined to be four within their
 scenario.
In the rest of this paper
 we consider
this interesting problem and we propose
an alternative idea to determine the dimensionality of
space time.
We first review Brandenberger and Vafa's discussion shortly and
give our reconsideration next.

In their scenario,
  an expanding universe begins  as a result of
 accidental growth of the torus universe along
some $d$ directions in nine dimensional space.
  In order for the accidental
perturbations to grow,
the strings should collide with
each other frequently so as to diminish the
 number of their winding modes along $d$-directions,
because the winding modes slow down the expansion.
If they intersect they would probably unwind.

On the other hand
 the string is a two
dimensional entity if
seen in space time.  They argued that in order for two
strings to collide with finite probability, $d+1$ must be smaller
than or equal to
$2+2=4$, otherwise  space time is too broad for two world sheet to have an
intersection.
This is their derivation of the relation $d+1\leq 4$.
They said that the universe continues to make trials and errors until they
learn that less than or equal to three spatial directions only become
 large.
When the universe finished the course,
 $d\leq 3 $ dimensional large universe would
have been born.  Thus, while they presented the intriguing idea for
understanding why $d\leq 3$, they did not succeed in giving
compelling reasons why $d$ should be $3$.

Now we reconsider their argument.
We can not fully agree with their discussion to
determine the dimensionality  because of the following reasons.
  First, it is true that in the
point particle case especially in $ \phi^4$ theory in $R^d$,
  the correlation
functions are known \cite{GJ}
 to be represented in terms of a random walks in $R^d$.
   From this, it is rigorously proven
 that the theory is free if $d+1\geq 5$.
However this is not the case for the string theory.
We have repeated  the
same analysis as $\phi^4$ theory
 in the light cone string field theory \cite{KK}.
We have found that
$\Phi^3$ interaction term  prevents us from constructing an analogous
 representation as the $\phi^4$ theory.

Second, it is true that the low energy point particles
cannot travel to the compact direction since the unit momenta are too
 heavy in such
direction.    This enforces the point particles effectively
 confined in the
 $d$ dimensional space.
In this case it is only $d$ directions
 that  the point particles can utilize to colide.
However, we are now concerned with the collision of the strings
 long enough to wind the torus universe many times.
One can imagine with ease that these strings
need not be
 confined in the $d$ dimensional space and can  move in any directions.
Then what aspects are relevant to the string case?

Let us recall what occurs to the point particles
in the expanding universe.  The expansion makes the mean separation of
particles larger.
If the time scale of expansion time get shorter
 than that of interaction time, the interaction
is effectively frozen.

This consideration gives rise to the following interesting possibility.
Let us assume that the accidental expansion in the $d(\leq 9)$  directions
is always too fast to keep  their mutual intersections.
If it is true, the expansion in the $d$ directions play a negative role
with respect to the unwinding of the winding strings.

To put it another way, the
 winding strings only can use the rest $9-d$ dimension to unwind.
  Hence the greater value of $9-d$ is favored from the view point of
killing the winding modes along $d$-directions.
This implies the inequality opposite in direction
from that of  Brandenberger and Vafa.
One may say that this implies that $d=1$ is preferred.  However
 the story is not so simple.

As we have observed in the expression (\ref{eq:entn}),  the entropy
produced in the process of the unwinding is proportional to the expanding
volume.     This means
 that the universe prefers to expand as many
dimensions as possible.  The larger value of $d$ is preferred from the
second law of thermodynamics, the entropy increase.
This effect will compete the preceding one.
   Our idea is that the number of
 the expanding space dimension is determined to be
three as a result of this competetion.
This idea is not yet formulated on a rigorous ground at present.
However we think it is one of the plausible candidates of a mechanism
to fix the space time dimension.  We are planning to
make a numerical simulation of the strings to obtain some indications
to this idea.

\acknowledgments

  The present author thanks all the researchers in the high energy
theory group and the group of the theoretical astronomy in Tokyo
Metropolitan University
for their continuous encouragements and valuable comments on this work.
The author especially thanks Dr.H.Minakata.   From the beginning of this
study he continuously gave the author valuable advices in many aspects
of the theory. This work was in part supported by Fund for Special
Research Project at Tokyo Metropolitan University.

\unletteredappendix{}

In this appendix we  examine what
 kind of correction is added when
we include the full quantum statistics.
  Rough estimation
of it has been done in \cite{BG}. We will consider this in our framework.
 As a result we can show that the correction  in our case is  very small.

It is known \cite{Deo} that the thermodynamical functions in the canonical
formalism and the micro canonical formalism is
connected through the Laplace transformation.  If we denote the single
canonical partition function for the non-winding
string as $\tilde f_N(\beta) $, this can be explicitly written as
\begin{equation}
\int_{-i\infty}^{i\infty}\bar d\beta
  \hbox{exp} \left( \tilde f_N(\beta) \right)
 \hbox{e}^{\beta \varepsilon}=
  \hat A(\varepsilon,v)\hbox{e}^{{\beta_H \varepsilon}}.
\label{eq:trans}
\end{equation}
within the M-B approximation,
where $\bar d\beta=d\beta/2\pi i$.

The multi string state density with all the statistical effects
$\omega_N(\varepsilon)$ is given by the similar integral, but the integrand
should be changed  to
\begin{equation}
Z(\beta)=\prod_{r=1,r:odd}^\infty
  \hbox{exp}\left[{1\over r}\tilde f(r\beta)\right].
\end{equation}
Namely the integrand is a product of the functions
${1\over r} \tilde f(r\beta) $.
The equation (\ref{eq:trans}) readily implies
\begin{equation}
\int_{-i\infty}^{i\infty}\bar d\beta
  \hbox{exp}\left({1\over r}
\tilde f_N(r\beta)\right)\hbox{e}^{\beta \varepsilon}=
{1\over r}\hat A(\varepsilon/r,v/r)\hbox{e}^{{\beta_H \varepsilon}/r}.
\end{equation}

Here we reminds the readers of a well known formula
 that the inverse Laplace transform of the product of
 two functions is equal to the convolution
 of the inverse Laplace transforms of the two functions.  Namely we have
\begin{eqnarray}
 \int_{-i\infty}^{i\infty}&\bar d&\beta \tilde \phi_1(\beta)
\tilde \phi_2(\beta)\hbox{e}^{\beta \varepsilon}\nonumber\\
&=&\int_0^\infty d\varepsilon_1d\varepsilon_2
 \phi_1(\varepsilon_1)\phi_2(\varepsilon_2)
  \delta(\varepsilon-\varepsilon_1-\varepsilon_2).\nonumber\\
\end{eqnarray}
Now the repeated use of this
  formula leads us to the
the multi state density with the full quantum statistical corrections:
\begin{eqnarray}\hat \omega_N(\varepsilon,v)
  =\lim_{k\rightarrow\infty}&\int_0^\infty &
d\varepsilon_1d\varepsilon_3\cdots d\varepsilon_k
    \nonumber\\
 &\times &\hat \alpha(\varepsilon_1,v)\hat \alpha(\varepsilon_3,v/3)\cdots
\hat \alpha(\varepsilon_k,v/k)\nonumber\\
 &\times&
\delta\left(\varepsilon-\sum_{r=1}^k r \varepsilon_r\right)\nonumber\\
\label{eq:aas}
\end{eqnarray}
where we have set $\alpha(\varepsilon,v)
=A(\varepsilon,v)\hbox{e}^{{\beta_H \varepsilon}}$.
All the summation and product indices in this appendix mean to run only odd
integers unless otherwise stated.

Now we are going to examine the size of the corrections to the
M-B approximation using (\ref{eq:aas}).
Because $ \alpha(\varepsilon,v) $ does
 not vanish only for $\varepsilon\geq m_1$,
the product of (\ref{eq:aas}) is actually a finite product.

For the range $k {m_1}\leq \varepsilon <(k+1){m_1}$
with $k$ being an integer we consider the
sequence of odd numbers such that
$1\leq r_0<r_1<\cdots<r_l$ and $r_0+r_1+\cdots+r_l=k$.  Only $r_0$ has the
possibility to become unity.  All the corrections acquired by
$\omega_N(\varepsilon,v)$ in this range are written as the form
\begin{eqnarray}
L=\int d\varepsilon_{r_0}\cdots d\varepsilon_{r_l}&{\alpha}
(\varepsilon_{r_0},V/r_0)
      \cdots {\alpha}(\varepsilon_{r_l},V/r_l)\nonumber\\
    &\times\delta(\varepsilon-(r_0\varepsilon_{r_0}+
\cdots+r_l\varepsilon_{r_l})).
         \nonumber\\
\label{eq:capl}
\end{eqnarray}
If we set ${v/\eta m_1^\eta}=w$ and $\varepsilon/{m_1}=x$,
${\alpha}$ is rewritten as
\begin{equation}
{\alpha}={1\over {m_1}}\hbox{exp}[h(x-w)+w+{\beta_H m_1} x].
\end{equation}
\widetext
Using it and making  a
 variable change $\varepsilon_{r_j}={m_1} x_j$ (
Here  $j$ runs even and odd integer.) enable
us to rewrite $L\hbox{e}^{-{\beta_H \varepsilon}}$ as

\begin{equation}
L\hbox{e}^{-{\beta_H \varepsilon}}={m_1}^l\int dx_0\cdots dx_l
   \hbox{exp}
    \left[ \sum_{j=0}^l ( h(x_j-w/r_j)+w/r_j-{\beta_H m_1}(r_j-1)x_j)\right]
   \delta\left(x-\sum_{j=0}^l r_j x_j\right).
\end{equation}
This is a typical form the  corrections added
 to $A(\varepsilon,v)
={\alpha}(\varepsilon,v)\hbox{e}^{-{\beta_H \varepsilon}}$.

\narrowtext
This function is expressed as an integration of a product of
functions of the form
\begin{equation}
\hbox{exp}\left[h(x_j-w/r_j)+w/r_j-{\beta_H m_1}(r_j-1)x_j \right].
\label{eq:exhw}
\end{equation}
First we consider the case $r_0=1$.
We already know that this peaks at
$x_0^{max}=w+c $ with the height
${1\over w^{1/\eta}}\hbox{exp}(w)$,
where $c $ is the point such that $h^\prime(c)=0$.

Next we consider the general case $r_j\not=1$.
In this case  the function is strongly damped
 by the exponential suppression $\hbox{exp}(-{\beta_H m_1}(r_j-1)x_j)$.
Actually the
position of the peak is now located around $x_j^{max}\sim 1$.
  It can be  verified by examining
where the
derivative of the exponent of (\ref{eq:exhw}) changes its sign from
positive to negative.
The derivative in question  is written as
$ h^\prime(x_j-w/r_j)-{\beta_H m_1} (r_j-1)$.  We recall here that
we  had the information on the derivative $h^\prime(x)$ because
it is essentially the microcanonical temperature examined before, see
(\ref{eq:microM}).
We know that
\begin{eqnarray}
h^\prime(x-w)=\partial_x \ln A&=&{m_1} \beta_N(\varepsilon,v)\nonumber\\
     &=&{\beta_H m_1}(\beta_N(\varepsilon,v)/{\beta_H }).\nonumber\\
\end{eqnarray}
{}From our previous numerical analysis (see Fig.\ref{fig7})
we know that $\beta_N/{\beta_H }$ is very
close to unity except $x\sim 1$, we observe that the above derivative is
negative all the range except for the very edge $x_j\sim 1$.
This means that the position of the peak is $x_j^{max}\sim 1$,
and accordingly
its height is negligibly small. It is no longer
the exponentially large like $\sim \hbox{exp}({v/\eta m_1^\eta})$.
This is always true if (\ref{eq:capl})
contains the factor with $r_j\not=1$.
Consequently we conclude that any
corrections to M-B approximation are very small.

\figure{The plot of the single string state density of the totally
compact model normalized
 as $f(\varepsilon)\hbox{e}^{-{\beta_H \varepsilon}}$. This
 is predicted to tend to $1/\varepsilon$. \label{fig1}}
\figure{The plot of  $\varepsilon
 f(\varepsilon)\hbox{e}^{-{\beta_H \varepsilon}}$. This shows the predicted
 asymptotic behavior.   \label{fig2}}
\figure{The semi-log plot of
 $f(\varepsilon)\hbox{e}^{-{\beta_H \varepsilon}}/V$ in the case that
the space having $D=3$ open dimensions.    \label{fig3}}
\figure{The semi-log
 plot of $\varepsilon^{5/2}f(\varepsilon)
\hbox{e}^{-{\beta_H \varepsilon}}/V$. This shows
the expected asymptotic behavior.   \label{fig4}}
\figure{
 The plots of $\ln A(\varepsilon,v)$
 versus $\varepsilon$ at growing values of $a$'s.
The figure (a) is for $a=1$, (b) is for, $a=1.4$, (c) is for
$a=1.6$, (d) is for $a=2.4$ and (e) is the close view around
$\varepsilon\sim m_1$ of (d). \label{fig5}}
\figure{
The plot of the favored energy of the non-winding strings
$e_N(a)$ normalized
by $a^3$ versus $a$.  This is consistent with the analytic estimate.
 \label{fig6}}
\figure{
The plot of $\ln A(e_N(a),v)/v$ versus $a$.  This is
 consistent with
 the analytic estimate. \label{fig7}}
\figure{
The plots of the micro canonical
temperature of the non-winding strings $\beta_N(\varepsilon,v)$
versus $\varepsilon$ at growing values of $a$'s.
The figure (a) is for $a=1$, (b) is for, $a=1.4$, (c) is for
$a=1.6$, (d) is for $a=2.4$ and (e) is the close view around
$\varepsilon\sim m_1$ of (d). \label{fig8}}

\end{document}